# Ferroelectric Domain and Switching Dynamics in Curved In$_2$Se$_3$: First Principle and Deep Learning Molecular Dynamics Simulations

[‡]Dongyu Bai[1], Yihan Nie[2,1,*], Jing Shang[3], Minghao Liu[1], Yang Yang[4], Haifei Zhan[2], Liangzhi Kou[1,*] and Yuantong Gu[1,*]

1. School of Mechanical, Medical and Process Engineering, Queensland University of Technology, Brisbane, QLD 4001, Australia

2. College of Civil Engineering and Architecture, Zhejiang University, Hangzhou 310058, China

3. School of Materials Science and Engineering, Shaanxi University of Science and Technology, Xi'an, 710021 China

4. State Key Laboratory for Mechanical Behavior of Materials, Xi'an Jiaotong University, Xi'an 710049, China

**Abstract**

Complex strain status can exist in 2D materials during their synthesis process, resulting in significant impacts on the physical and chemical properties. Despite their prevalence in experiments, their influence on the material properties and the corresponding mechanism are often understudied due to the lack of effective simulation methods. In this work, we investigated the effects of bending, rippling, and bubbling on the ferroelectric domains in $In_2Se_3$ monolayer by density functional theory (DFT) and deep learning molecular dynamics (DLMD) simulations. The analysis of the tube model shows that bending deformation imparts asymmetry into the system, and the polarization direction tends to orient towards the tensile side, which has a lower energy state than the opposite polarization direction. The energy barrier for polarization switching can be reduced by compressive strain according DFT results. The dynamics of the polarization switching is investigated by the DLMD simulations. The influence of curvature and temperature on the switching time follows the Arrhenius-style function. For the complex strain status in the rippling and bubbling model, the lifetime of the local transient polarization is analyzed by the autocorrelation function, and the size of the stable polarization domain is identified. Local curvature and temperature can influence the local polarization dynamics following the proposed Arrhenius-style equation. Through cross-scale simulations, this study demonstrates the capability of deep-learning potentials in simulating polarization for ferroelectric materials. It further reveals the potential to manipulate local polarization in ferroelectric materials through strain engineering.

**Keywords:** 2D ferroelectric, α-$In_2Se_3$, polarization switching, deep learning potential, strain engineering

Over the past two decades, 2D materials have emerged as a new research hotspot, due to their unique physical and chemical properties, as well as promising applications in biosensors,[1, 2] solar photovoltaics,[3] catalysis,[4-6] electronic devices,[7-9] and more. The atomic-thick structure distinguishes them from their bulk counterparts in terms of mechanical properties, thermal and electrical conductivity. However, the weaker bending stiffness compared to their in-plane modulus[10] makes it easy to induce out-of-plane bending and crumpling, especially when influenced by the substrate,[11-13] interface liquid,[14, 15] and thermal strain.[16-18] In fact, the inevitable wrinkled structures have been commonly observed in graphene,[19] transition metal dichalcogenides,[20] black phosphorene, and many others. The corrugations have been shown to significantly affect the electronic, mechanical, electrical, and optical properties of 2D materials, thereby impacting the performance of associated devices.[21, 22] For example, the crest of graphene's nano-bubbles or waves exhibits enhanced catalytic activity for HER or ORR[23, 24] than the area of flat graphene. The rippling deformation in $MoS_2$ is found to have a detrimental effect on electron mobility, removing them could result in reaching ultimate limits of mobility in this material to achieve optimal performance of electronic device. Meanwhile, through proper strain engineering, the out-of-plane deformation can be used for optimizing performance or designing novel functional devices.[25, 26]

2D ferroelectrics, such as $In_2Se_3$,[27-29] $MoTe_2$,[30] $CuInP_2S_6$,[31-33] and the SnTe family,[34-36] are a unique class of van der Waals materials that exhibit stable spontaneous and switchable polarization under external stimuli. Mechanical bending or rippling deformation can well couple with electric polarization due to the generated interlayer movement or ion migration,[27, 37] leading to flexoelectric phenomena or ferroic domains that enable the measurement of strain gradient in mechanical structures electrically. Rippling/bending deformations therefore gain the ability to weaken, strengthen, or reverse polarization. For example, polarization vortex or polar skyrmions were realized at room temperature in ferroelectric bubbling domains induced by lattice mismatch strain in $(PbTiO_3)_n/(SrTiO_3)_n$ heterostructures.[38, 39] Ripple-induced ferroic phase transition and domain switching have been observed in 2D ferroelectric GeSe layers, indicating their potential application in flexible electronics.[40] Despite their fundamental and practical importance, a deep understanding at the atomic level for the effect of wrinkles on polarization switching, especially in terms of out-of-plane ferroelectricity, remains scarce due to the lack of appropriate simulation methods and the requirements on extreme-high resolution of experimental measurements.

In this work, we studied the impacts of bending, rippling and bubbling deformations of α-$In_2Se_3$ on its out-of-plane ferroelectric polarization direction, using the joint Density Functional Theory (DFT) calculations and Molecular Dynamics (MD) simulations. The details of the methodology used in the work can be found in Supporting Information & Figure S1. DFT calculations found that the energy barrier of ferroelectric switching is positively correlated to uniaxial tensile strain, the reduced energy barrier can potentially enable polarization switching to the lower energy state generated by bending asymmetry. Such automatic polarization reversal is observed in $In_2Se_3$ nanotube. To further explore the switching dynamics in complex strain deformation, deep-learning potentials[41-43] are developed and used in larger scale MD simulations. The effects of curvature, strain gradient and temperature on polarization switching are comprehensively investigated. It is found that ferroelectric domains can be created and switched by the rippling and bubbling deformations, and the switching time can be controlled by curvature and temperature. Our results provide a new reference for the strain engineering of two-dimensional ferroelectric materials as flexible electronic devices.

## Switching Energy Barrier Influenced by Strain

The spontaneous polarization in α-$In_2Se_3$ results from structural asymmetry along the out-of-plane direction, where the position of the middle Se layer plays a critical role. The middle Se layer can be shifted along in-plane direction with two steps, causing the polarization reverse (Supporting Information, Figure S2b). The polarization stability is protected by the energy barriers in the shifting process. Such energy barrier can be affected by strain, which is first investigated through uniform tensile strain analysis, providing fundamental knowledge for later analysis of complex strain status.

According to DFT calculation, the in-plane tensile strain can significantly increase the energy barrier. Such barrier is increased by ~300% (**Figure 1a**) with 4% tensile strain. In contrast, it is significantly reduced by the compression strain, which is close to 0 with -3% strain. The results imply that the polarization is easier to switch under the compression strain state, while difficult for tensile state. The phenomena can be intuitively understood from the structural deformation, where the thickness along out-of-plane direction will expand under in-plane compression deformations, giving more space for the motion of In and Se layers in the out-of-plane direction, which leads to a smaller switching energy barrier. This can also be seen from electron localization function (ELF) along the (1 0 0) plane (Figure 1c). Taking the P↑-$In_2Se_3$ monolayer as an example, more localization of electron is observed in tensile case (+3%) than that in compressive model (-3%). Thus, the tensile state is less prone to switch polarization direction than the compressed state.

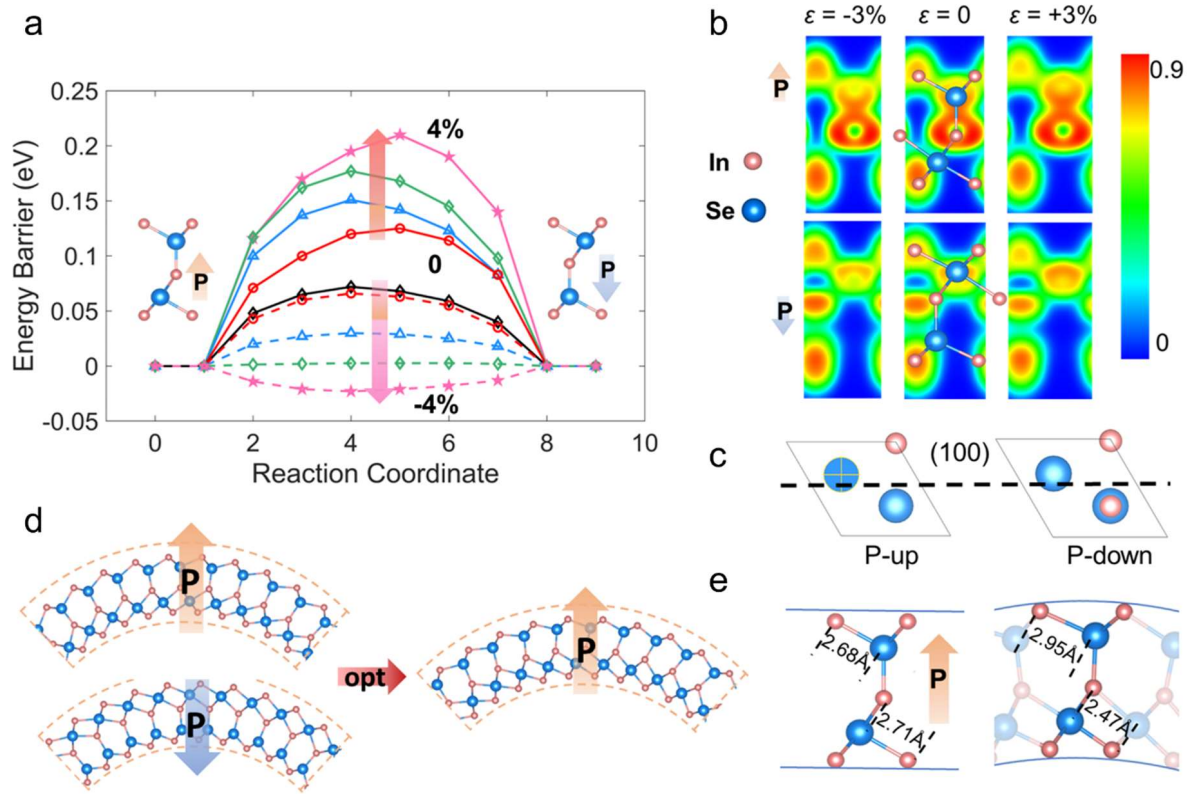

**Figure 1.** (a) Energy barriers for polarization switching directions of monolayer α-In$_2$Se$_3$ under various in-plane strains. (b) The electron localization function (ELF) maps under various in-plane strains. The upper and lower panels present the P↑ and P↓ monolayer In$_2$Se$_3$ respectively. (c) The cross section corresponding to ELF maps of P↑ and P↓. (d) The initial and optimized partial structure of In$_2$Se$_3$ nanotube from P↑ and P↓ states. (e) The variation in In-Se bond length before and after bending a nanotube structure with a radius of 22.60 Å.

To verify the different impacts of strain on the phase transition, we artificially build a In$_2$Se$_3$ nanotube with the radius of 22.6 Å to investigate the bending effect. Depending on the wrap direction, two models with initially different polarization states are constructed (Figure 1d). After the structural optimization, both structures transform into the same configuration, i.e., the polarization pointing towards outside as shown in Figure 1d. As the bending deformation introduces asymmetry into the systems, resulting in two distinct polarization directions with different energy states, wherein the outward direction holds a lower energy value. The In-Se bond lengths are stretched to 2.95 Å at the outer layer (originally 2.68 Å), but compressed to 2.47 Å (originally 2.71 Å) at inner layer (Figure 1e). The outside of the tube is subjecting the tensile strain while the inner side is subjecting the compressive strain. In this case, the middle Se layer tends to move to the compression region, as the space in the out-of-plane direction is expanded. Therefore, regardless of the initial polarization direction and structural models, the polarization direction in In$_2$Se$_3$ tubes always points outwards after the structural optimization, and the middle Se layer prefers to migrate to the compression side under bending conditions when both tensile and compressive strains coexist. Limited by the computational capacity of DFT calculation, the influence of bending curvature and complex strain deformation will be investigated in the following MD simulations with deep learning potential.

## Validation of Deep Learning Potential

To conduct MD simulations for complex systems, we developed the force fields of In$_2$Se$_3$ with the DeepMD-kit.[41] The training database is based on open-source data from the recent work by Wu et al,[44] which was generated by DP-GEN method and contained 22600 monolayer structures such as $\alpha$, $\beta$, $\beta'$, $\beta'_1$, and $\beta'_2$ phases and 2163 bulk structures. More details of the potential training including the descriptor, training neuron layers and accuracy can be found from the Supporting information 1. Our benchmark calculations indicate that the potential performs remarkably well in both energy and force predictions. Across nearly 8,500 examples, the mean absolute error (MAE) is around 3.9 meV/atom (Supporting Information Figure S3). Significantly, the deep learning potential can accurately predict the energy barrier under various strain conditions and phase transitions, demonstrating remarkable consistency with previous DFT results (Figures S2，S3 & S4). All MD simulations in the following will be based on this deep learning-force field and implemented in LAMMPS software.

## Switching Dynamics under Pure Bending

Although DFT simulations can predict the polarization reversal behaviors, it doesn't consider thermal disturbance from the environments. Based on the trained force field, we are able to study the switching dynamics and complex systems at various temperatures. Here the $In_2Se_3$ nanotube with radius of 28.25 Å are selected and built. Consistent with the findings from DFT simulations, the polarization for the configuration which initially points inside is reversed after the structural relaxation, with a sudden energy drop, as seen in **Figure 2a** and Figure S5. The switching time $\tau$ (defined as the time for the total energy reduces to the average energy of the two states) is 17 ps at 370 K (Figure 2a), and reduces to 10 ps when the temperature was increased to 380 K, the polarization switching process is accelerated by the increased thermal disturbance. The underlying mechanism can be attributed to increased kinetic energy of the atoms, which enhances the likelihood of overcoming the energy barrier and accelerates ferroelectric switching behaviours. The comprehensive simulations show the evolution of the switching time $\tau$ of the tube exhibits an exponential relationship with temperature, see Figure 2c. The switching process can be approximate to chemical reactions, where the energy barrier corresponds to the activation energy. Consequently, the switching time should be the reciprocal of the reaction rate, which could be effectively modeled using an Arrhenius-style equation:

$$\tau = Ae^{\left(\frac{E_b}{k_bT}\right)} + t_m \qquad (1)$$

where $A$ is the amplitude fitted as $9.57\times10^{-12}$ ps, $E_b$ is the energy barrier fitted as 0.895eV and constant $t_m$ equals 1.90 ps, $k_b$ is the Boltzmann constant, and $T$ is the temperature. The proposed equation can well fit the function between switching time and temperature, which further proves the existence of energy barrier during the switching process. The constant $t_m$ indicates that for a certain curvature, there is a minimal time required for the switching process.

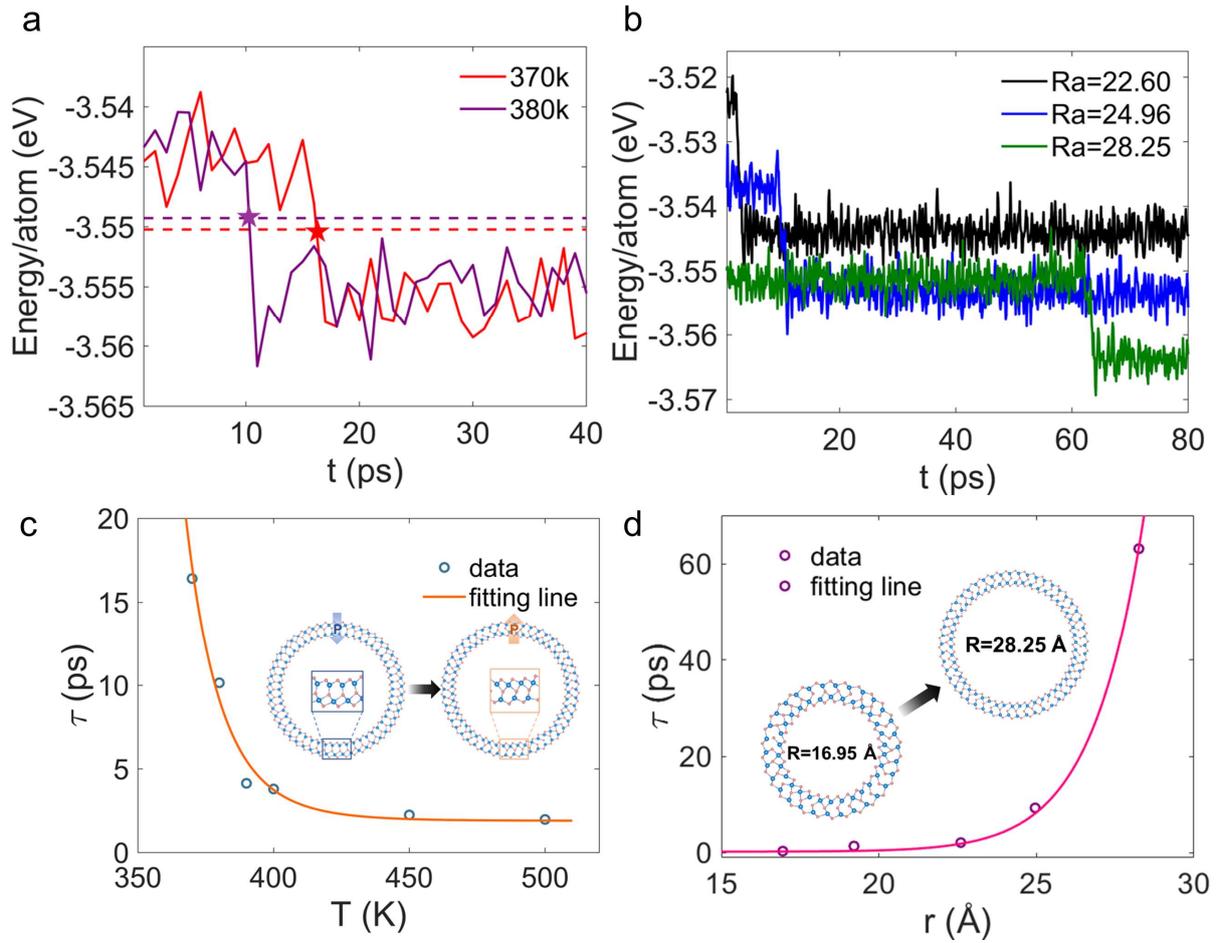

**Figure 2.** DLMD results for In$_2$Se$_3$ tubes. (a) Energy history for a tube with radius 28.25 Å at different temperatures, represented by the solid line. The dashed line is the average value between the two polarization states. The point of intersection is selected as the switching time. (b) Energy history with time for a group of tubes with radius from 22.60 Å to 28.25 Å at 300 K. (c) The change of switching time $\tau$ with temperature and its fitting curve of tube with radius 28.25 Å. The illustration is a comparison of the configurations before and after the polarization switching. (d) The change of switching time $\tau$ with different radii and its fitting curve of tube at 300 K. The illustration is the configurations of tubes with a radius of 16.95 Å and 28.25 Å after the polarization switching.

The switching dynamics is also affected by the bending curvature, as the energy barrier can be influenced by the strain. The switching time $\tau$ for nanotubes with radius of 22.6 Å, 24.96 Å, and 28.25 Å are simulated under 300 K (Figure 2b). The polarization reversal takes 3 ps for the smallest tube (R=22.6 Å), but much longer for the other two (10 ps for R=24.96 Å and 63 ps for R=28.25 Å respectively). The trend remains true for even smaller tubes (Supporting Information, Figure S6). The findings can be rationalized by our DFT results, the increasing of the bending strain exacerbates the asymmetry, leading to a larger energy variance between two polarization state, see Figure 2b. On the other hand, the energy barrier is also influenced by strain, which is quantified through previous Arrhenius style equation according to the switching time. The comprehensive investigation can fit an Arrhenius style equation about the switching time ($\tau$) as a functional of the radius:

$$\tau = A\, e^{\frac{l_e \sqrt{r}}{k_b T}} + t_m \tag{2}$$

where the amplitude $A$ equals $6.245 \times 10^{-14}$ ps, $l_e$ is the coefficient for energy barrier 0.168 eV/Å$^{\frac{1}{2}}$, $r$ is radius of nanotube model, and constant $t_m$ equals 0.25 ps. $k_b$ is the Boltzmann constant, and $T$ is temperature. The function can well describe the relation between switching time and curvature (Figure 2d). The fitting equation indicates a linear relationship between the energy barrier and the 1/2 power of the radius of curvature. The constant $t_m$ indicates a minimal time for switching process at 300 K. Both models (equation 1 and 2) with separate fitting processes give the similar energy barrier prediction values, i.e., 0.89 eV, for 28.25 Å nanotube, proving that the parameters $E_b$ and $l_e$ can present the energy barrier value, which is not just a mathematical fitting parameter.

## Switching Dynamics with Rippling Deformation and the Lifetime of the Transient Polarization

Besides the pure bending, rippling deformation is more commonly observed in 2D materials when experiencing compression forces at the boundary. Here, a mono-sinusoidal shaped model is created to mimic rippling deformation. Additional displacement is added in out-of-plane direction, the coordinates of the neutral layer follows the formula $\Delta z = m \times sin(\frac{2\pi}{\sqrt{3}l_a \times n_x}x)$, where $x$ is the position in [1 0 0] direction, and $\Delta z$ are the preset displacement of each atom in [0 0 1] direction according to the function of the neutral layer, $m$ is the rippling amplitude, $l_a$ is the periodic length along the $x$ direction, and $n_x$ is the lattice number in one period. The simulations were conducted at different temperatures (ranging from 300 K to 670 K) and with different rippling amplitudes ($m$ =15 and $m = 20$). Due to the opposite curvature at the crest and trough, the local polarization has opposite preference direction in the two regions, resulting in localized ferroelectric domains. After the structural relaxation, a prominent phase transition occurs in the left region, which is situated in the wave trough with a positive curvature. This phase transition is caused by the in-plane shift of the middle Se layer, resulting in a polarization direction switch from [0 0 1] to [0 0 -1], similar to the phenomenon observed in nanotubes (Figure 2). In contrast, the polarization in the right region, located at the wave crest with a negative strain curvature, remains unchanged throughout the process. These structural phase transitions and localized polarization reversals are also observed in other models with different rippling amplitudes ($m$=15 Å) and temperatures. Meanwhile, larger rippling amplitudes accelerate the polarization reversal at the wave trough. In the case of the model with $m$=15 Å under a temperature of 300 K, the switching time predicted by equation 2 will be at least $4.41 \times 10^3$ s, but it can be accomplished when the temperature is increased to 550 K. Furthermore, when the amplitude is increased to 20 Å (at T = 550 K), the phase transition time is reduced to 30 ps (as shown in the Supporting Information, Figure S7).

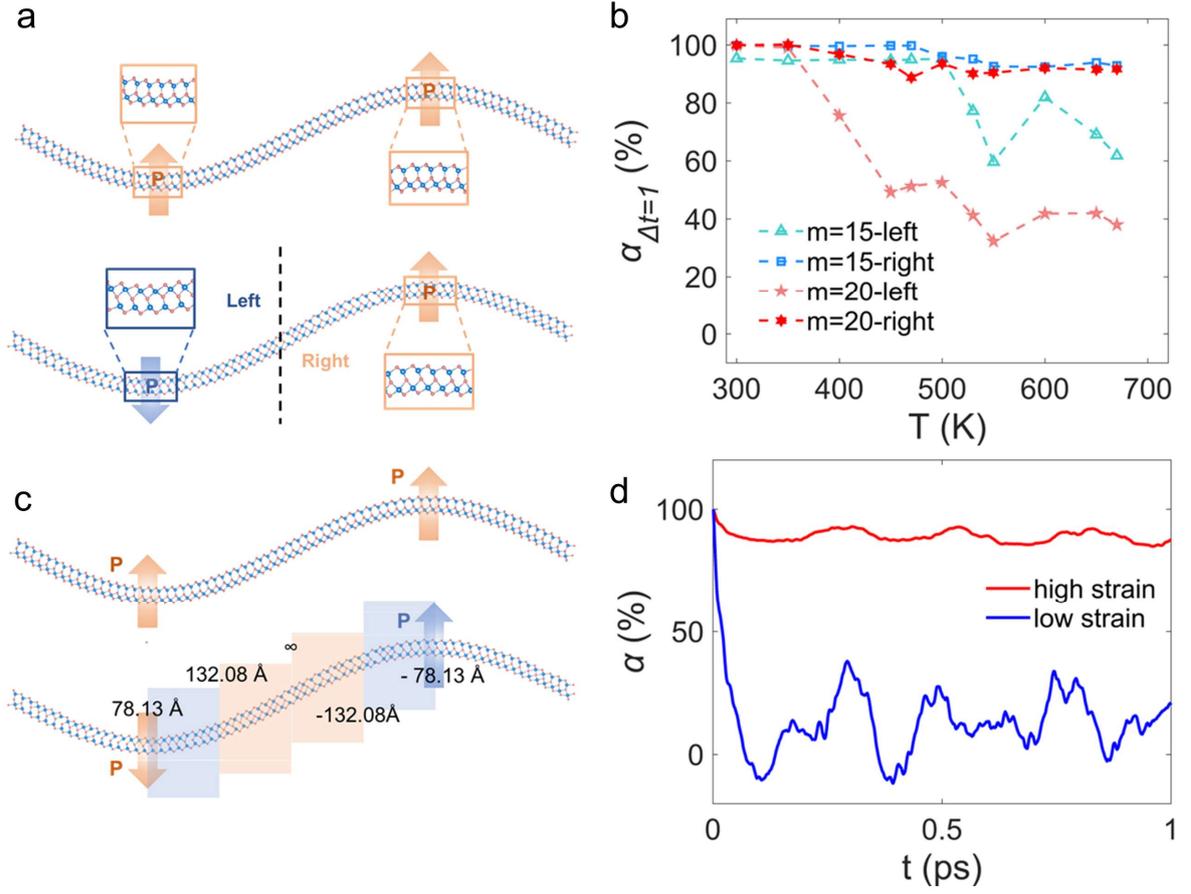

**Figure 3.** DLMD calculation results in rippling models. (a) The monolayer rippling configuration with $m$=20 at 450 K. The top image represents the structure at 0 ps, and the bottom image represents the structure at 85 ps. (b) The correlation of polarization of the left and right regions at 1 ps as a function of temperature. (c) Division of the rippling model into high strain and low strain regions, with corresponding curvature radii indicated. (d) Temporal variation of polarization correlation in different strain regions.

Due to the minimal bending strain in the transition region, the energy of two polarization states is similar. Thus, the direction preference of polarization is not obvious and can switch frequently. To quantify the lifetime of such transient polarization, we investigate the local polarization lifetime by tracking local momentum auto-correlation functions.

$$\alpha(t) = \frac{<\int \vec{D_i}(t_0)\vec{D_i}(t_0+t)dt_0>}{\alpha_i(0)} \tag{3}$$

to describe the difference of the polarization at the time interval of $t$. $\vec{D_i}(t_0)$ is the value of the dipole moment of chunk $i$ at $t_0$. $t$ is the time interval. The autocorrelation function starts at $t=0$, $\alpha_i(0)$ normalize the starting point to 1 for comparison. After the polarization switching, a lower correlation between polarization directions indicates a shorter lifetime for polarization. To facilitate a clear and intuitive comparison of the impact of curvature on the polarization direction, the model is divided into distinct left and right chunks based on regional polarization directions (**Figure 3a**).

The investigation reveals that the autocorrelation function of the left region experiences a more rapid decay, in contrast to the nearly unaltered value observed for the right one. This is due to the preferred polarization direction of the left region opposes the preset orientation, leading to a switching occurring. As the autocorrelation function is influenced by the switching time within the left region, its autocorrelation function is smaller than that of the right region (Figure 3b). Using the model at 450 K as an illustration, the switching time $\tau$ is 70 ps. The radius of curvature at the crest of the model is 78.13 Å, incorporated into equation 2, the calculated switching time significantly surpassing the observed switching time. This discrepancy can be ascribed to the complexity of the rippling model, which encompasses a bending state accompanied by compression. These factors lead to a high initial state and a reduction in the energy barrier, thereby accelerating the switching of the model. When amplitude increases, the curvature increases, leading to a shorter switching time $\tau$ and more rapid decay of autocorrelation function in the left region (Figure S8).

To show the stability of the polarization at small curvature, we divided the rippling model with $m=20$ Å into different regions after relaxation, based on the curvature radius (Figure 3c). In the high strain region, the curvature radii range from 78.13 Å to 132.08 Å, while the radii in the low strain region are above 132.08 Å. The polarization autocorrelation functions at 450 K show different characteristics in the high and low strain regions (Figure 3d). The autocorrelation function in the low strain region drops significantly, while the high strain region remains above 80% after 1 ps. Furthermore, most structures in the low strain region maintain the $\beta$ state during relaxation (Figure 3c). In the areas with high bending strain regions, the energy difference is more pronounced, leading to enhanced polarization direction stability. Conversely, regions with low bending strain have small energy differences due to the increased symmetry. Thus, these regions exhibit more frequent polarization switching, resulting in a shorter lifetime of polarization state.

## Switching Dynamics with Bubbling Deformation and the Ferroelectric Domain

To further identify the combination effect of the curvature and tensile deformation on the polarization, we created the dual sinusoidal model to mimic the bubbling deformation. The model surface follows the equation $z = m \times sin\left(\frac{2\pi}{\sqrt{3}l_a \times n_x}x\right) \times sin\left(\frac{2\pi}{l_b \times n_y}y\right)$, where $m$ is the bubbling amplitude, more details about the models and calculation methods can be found from Supporting Information 2.

The surface is divided into small chunks and the transient local polarization distribution at 450 K at presented in **Figure 4a**. Compared with the rippling model, the domain area with significant bending strain shrinks. Consequently, most of the area experiences minor bending deformation, displaying little polarization direction bias. As a result, the transient local polarization lacks uniformity. According to previous simulations, the larger amplitudes would induce greater bending and tensile strains, resulting in a more stable polarization direction in time. However, when amplitude increases, the size of the domain exhibiting stable polarization direction shrinks. Such phenomenon can be attributed to the competing influences of bending and stretching deformations. With increased amplitude, the curvature increment becomes less significant compared to the tensile strain's impact. Thus, the region dominated by bending strain shrinks and excessive stretching destroys the original polarization structure, causing the reduction of the area with stable polarization.

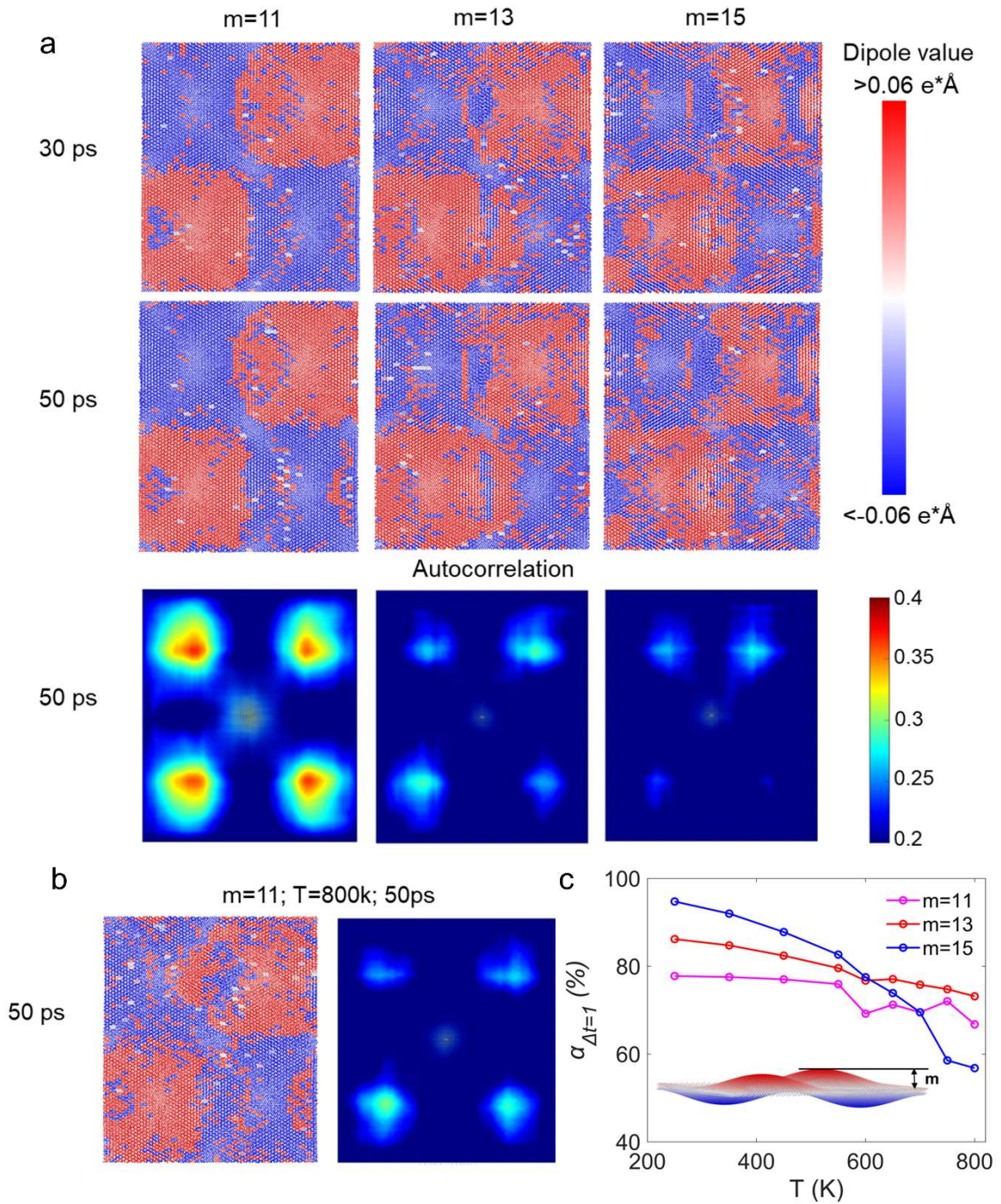

**Figure 4.** (a) The polarization distribution during relaxation and autocorrelation maps at $m$=11, 13 and 15 at 450 K in bubbling model. (b) The polarization distribution at 50 ps and autocorrelation map at $m$=11 at 800 K in bubbling model. (c) Structural stability $\alpha_{\Delta t=}$ of different $m$ with temperature and its trend line in bubbling model.

To highlight the region with stable polarization, image autocorrelation analysis is employed. The resulting autocorrelation image for the polarized region is displayed at the bottom of Figure 4a, where red regions indicate a higher degree of self-similarity and stronger correlation, corresponding to the areas where the pattern repeats, i.e., the the domain size of stable polarization region is larger. As shown in Figure 4a, the increase of amplitude can result in a decrease in the domain size of stable polarization region.

The study at various temperatures reveals a substantial influence of temperature on ferroelectric domains, wherein higher temperatures notably diminish the size of the stable polarization domain (Figure 4b and Figure S9). For instance, at 800 K, the stable polarization region is disrupted after 50 ps structural relaxation, leading to a homogenous mixture and of different polarization directions, presenting the ferroelectricity of $In_2Se_3$ disappeared over curie temperature.[45]

The influence of temperature and bubbling amplitude on the ferroelectric domain can be more clearly seen when we specifically exclude the impact of tensile effects. Here, the autocorrelation function is utilized to analyze the polarization lifetime within the high-strain region of the bubbling model (Figure S10, Figure S11 & Figure 4c). Under the same strain condition, $\alpha_{\Delta t=1}$ exhibits lower values at higher temperatures, indicating that local polarization stability decreases when temperature increases. Concurrently, the bubbling amplitude demonstrates a positive impact on stability under consistent temperature conditions. An increased strain gradient (corresponding to an increased bubbling amplitude value) leads to a higher $\alpha_{\Delta t=1}$, namely the longer lifetime in the polarization domain. For example, the value is 77% when $m$=11 Å at 250 K, but increases to 86% and 94% as the amplitude $m$ = 13 Å and

15 Å respectively. These results confirm that in the large bubbling model of $In_2Se_3$, previous relation between polarization stability and curvature still remains.

In summary, we report the polarization switching and its dynamics in strained monolayer $In_2Se_3$ through the joint DFT and DLMD simulations. We confirmed that the ferroelectric $In_2Se_3$ monolayer will undergo an automatic polarization reversal under the pure bending, the switching time is dependent on the environmental temperature and bending curvature. Research conducted using complex rippling and bubbling models reveals that lower temperatures and elevated bending strain foster larger domain sizes and longer lifetimes. Nevertheless, the tensile effects brought about by strain amplitude can lead to a contraction of polarization domain size within the complex model. Our research offers a promising approach for investigating strain effect on polarization dynamics of ferroelectric materials by conducting cross-scale simulations, and it provides valuable insights for the controlling polarization characteristics of ferroelectric materials by strain engineering in device fabrication.

## ASSOCIATED CONTENT

**Supporting Information**

The Supporting Information are available free of charge at .

Details for DL potential training, methods of model construction, computational details of DFT and MD simulations, validation results for DL potential, additional results for nanotube, rippling and bubbling models (PDF)

## AUTHOR INFORMATION

**Corresponding Authors**

**Yihan Nie** – *College of Civil Engineering and Architecture, Zhejiang University, Hangzhou 310058, China; School of Mechanical, Medical and Process Engineering, Queensland University of Technology, Garden Point Campus, Brisbane, QLD 4001, Australia;* Email: Yihan.nie@zju.edu.cn

**Yuantong Gu** – *School of Mechanical, Medical and Process Engineering, Queensland University of Technology, Garden Point Campus, Brisbane, QLD 4001, Australia;* Email: yuantong.gu@qut.edu.au

**Liangzhi Kou** – *School of Mechanical, Medical and Process Engineering, Queensland University of Technology, Garden Point Campus, Brisbane, QLD 4001, Australia;* Email: Liangzhi.kou@qut.edu.au

**Authors**

**Dongyu Bai** – *School of Mechanical, Medical and Process Engineering, Queensland University of Technology, Garden Point Campus, Brisbane, QLD 4001, Australia*

**Jing Shang** – *School of Materials Science and Engineering, Shaanxi University of Science and Technology, Xi'an, 710021 China*

**Minghao Liu** – *School of Mechanical, Medical and Process Engineering, Queensland University of Technology, Garden Point Campus, Brisbane, QLD 4001, Australia*

**Yang Yang** – *State Key Laboratory for Mechanical Behavior of Materials, Xi'an Jiaotong University, Xi'an, 710049 China*

**Haifei Zhan** – *College of Civil Engineering and Architecture, Zhejiang University, Hangzhou 310058, China*

**Author Contributions**

L.K. conceived the idea and supervised the research with Y.G. and Y.N.. D.B. and Y.N. jointly designed the simulation process and models. D.B. trained the DLMD potential and conducted the MD simulations, assisted by Y.Y. and H.Z.. J.S. and M.L. executed the DFT simulations. D.B. analyzed the data and composed the manuscript. All authors have given approval to the final version of the manuscript.

## ACKNOWLEDGMENTS

This work was supported by the Australian Research Council (Grant IC190100020 and DP200102546), the National Natural Science Foundation of China (Grant 12202254), and the High-performance Computing (HPC) resources provided by Queensland University of Technology (QUT). This research was undertaken with assistance of resources and services from the National Computational Infrastructure (NCI), which is supported by the Australian Government.

## ABBREVIATIONS

HER, hydrogen evolution reaction; OER, oxygen evolution reaction; DFT, density functional theory; DLMD, deep learning molecular dynamics;